\documentclass[12pt]{article}
\usepackage{float} 
\usepackage{amsmath}
\usepackage{slashed}
\usepackage{amsfonts}   
\usepackage{amssymb}

\begin{document}
\date{}
\title{{\bf{\Large Probing $ \eta $ deformed backgrounds with $ Dp $ branes}}}
\author{
 {\bf {\normalsize Dibakar Roychowdhury}$
$\thanks{E-mail:  dibakarphys@gmail.com, Dibakar.RoyChowdhury@swansea.ac.uk}}\\
 {\normalsize  Department of Physics, Swansea University,}\\
  {\normalsize Singleton Park, Swansea SA2 8PP, United Kingdom}
\\[0.3cm]
}

\maketitle
\begin{abstract}
In this Letter, based on the notion of Gauge/Gravity duality we explore the low frequency behaviour associated with the retarded two point correlators in the ground state of the strongly correlated quantum liquid that is dual to $ \eta $- deformed background in ($ 2+1 $)D. The massless charge carriers in the dual gauge theory are sourced due to some \textit{probe} $ N_f $ flavour $ Dp $ brane configurations in the bulk. In our analysis we stick to the NS sector and compute the two point correlators by turning on fluctuations associated with the worldvolume gauge fields in the bulk spacetime. Our analysis reveals the existence of holographic zero sound modes for ($ 1+1 $)D QFTs those are dual to bosonic $ \eta $ deformed $ AdS_3 \times S^{3} $ with vanishing RR fields.
\end{abstract}
\section{Overview and Motivation}
 $ AdS_5 \times S^{5} $ superstrings are known to be classically integrable for a long period of time \cite{Arutyunov:2009ga}-\cite{Hofman:2006xt}. The classical integrability of the superstring sigma model implies the existence of an infinite tower of conserved charges associated with the $ 2D $ superconformal field theory with the target space being the $ \frac{PSU(2,2|4)}{SO(4,1)\times SO(5)} $ supercoset \cite{Arutyunov:2009ga},\cite{Metsaev:1998it}. Given the integrable structure of $ AdS_5 \times S^{5} $ superstrings, the quest for the corresponding integrable deformation of the theory has been an active area of research for the last couple of decades\footnote{See \cite{vanTongeren:2013gva} for a nice and comprehensive review on this subject.} \cite{Lunin:2005jy}-\cite{Delduc:2013qra}. 
 
Very recently,  based on the general notion of classical Yang-Baxter (YB) deformations, the authors in \cite{Delduc:2013qra} had proposed an interesting as well as elegant way of deforming the $ AdS_5 \times S^{5} $ superstring sigma model which naturally breaks the supersymmetries of the original string sigma model still allowing the possibility to solve the system exactly. It turns out that at the classical level this deformation possesses two basic characteristic features of the original GS superstrings namely, (1) the existence of the fermionic kappa symmetry and (2) the Lax connection which thereby guarantees the existence of an infinite tower of conserved charges associated with the $ 2D $ sigma model. 

Soon after the discovery, various crucial features associated with the $ \eta $-deformed sigma model has been explored both from the perspective of the $ 2D $ worldsheet theory \cite{Delduc:2014kha}-\cite{Lunin:2014tsa} as well as that of the (deformed) target space geometry \cite{Kameyama:2014vma}-\cite{Ahn:2016egk}. However, the later direction of understanding becomes crucial as soon as one starts exploring various physical properties associated with the dual gauge theory at strong coupling. 

Considering these facts, the purpose of the present article is to perform a systematic analysis along this second direction with a view towards understanding the various physical properties associated with the dual gauge theory at finite charge density and/or chemical potential ($ \mu $). In particular, in the present analysis, we probe the dual gauge theory in its ground state and explore the \textit{low} frequency properties associated with the retarded two point correlators at strong coupling. The key idea is to obtain the dispersion relation looking at the pole of the retarded two point correlators in order to find traces of the zero sound mode \cite{Karch:2009zz}-\cite{DiNunno:2014bxa}. Typically, in a generic holographic Landau Fermi Liquid (LFL) theory, the leading order contribution to the dispersion relation goes linearly with the power of the corresponding momentum ($ \mathfrak{q} $) coupled to a real coefficient. The next subleading correction appears with a complex coefficient and goes quadratically with the momentum ($ \mathfrak{q}^{2} $) and which is thereby suppressed (compared to that with the leading term) in the limit of low momentum \cite{Karch:2009zz}.

 In order to compute the two point correlator, we \textit{probe} the $ \eta $- deformed target space with $ N_{f} (\ll N)$ flavour $ Dp $ branes and turn on fluctuations associated with the world volume gauge fields\footnote{For the purpose of our present analysis one might ignore the RR fields as well as the dilaton \cite{Edalati:2013tma} and consider the so called ABF background \cite{Arutyunov:2013ega} together with the NS-NS fluxes. } corresponding to these flavour $ Dp $ brane configuration.  In the language of $ AdS/CFT $ duality, this corresponds to turning on a global (diagonal) $ U(1)\subset U(N_{f}) $ operator and an associated chemical potential ($ \mu $) for the dual quantum liquid at strong coupling. At this stage, it is worth pointing out that in the present analysis we model only massless charged carriers in the dual gauge theory and we leave the corresponding scenario associated with the massive charge carriers for the purpose of future analysis. 

The organisation of the paper is the following: In Section 2, we start our analysis with a formal introduction to the $ \eta $- deformed background (associated with the original bosonic supercoset model \cite{Arutyunov:2013ega}) in $ 6 $D which could be realised as a consistent truncation of the original $ 10 $D background with a vanishing $ B $ field \cite{Hoare:2014pna}. We express the $ AdS_{3} $ sector of the $ \eta $- deformed target space in the so called Poincare patch with proper Lorentzian signature. As a next step of our analysis, in Section 3, we consider flavour $ Dp $ brane embeddings in this deformed background and compute the retarded two point correlators in order to explore the low frequency behaviour associated with the pole of the retarded correlator which eventually leads us towards the so called dispersion relation exhibiting a strong evidence in favour of the zero sound modes associated with the ground state of the system. Finally, we conclude our analysis in Section 4.  

\section{The background}
We start our analysis with a formal introduction to the $ \eta $- deformed $ AdS_3 \times S^{3} $ background which acts as a dual target space for the strongly correlated quantum liquid under investigation. In our analysis, we would be solely concerned with the bosonic sector of the full $ 10D $ solution \cite{Hoare:2014pna}. Under such circumstances, one could visualize the $ \eta$- deformed $ AdS_3 \times S^{3} $ supercoset as a consistent $ 6D $ reduction of the full $ 10D $ solution with a vanishing $ B $ field. The resulting background could be formally expressed as a direct sum of the individual spaces namely \cite{Hoare:2014pna},
\begin{eqnarray}
ds^{2}_{AdS_3 \times S^{3}}&=&ds^{2}_{AdS_3}\bigoplus ds^{2}_{S^{3}}\nonumber\\
&=&\left[ -\mathfrak{h}(\varrho)dt^{2}+\mathfrak{f}(\varrho)d\varrho^{2}+\varrho^{2}d\psi^{2}\right] \bigoplus\left[ \tilde{\mathfrak{h}}(\theta)d\varphi^{2}+\tilde{\mathfrak{f}}(\theta)d\theta^{2}+\cos^{2}\theta d\phi^{2}\right] 
\label{E1}
\end{eqnarray}
where, the individual metric functions could be formally expressed as\footnote{Notice that, here the deformation parameter $ \kappa $ is related to the original deformation parameter $ \eta $ as, $ \kappa = \frac{2\eta}{1-\eta^{2}}$ \cite{Arutyunov:2013ega}. Henceforth, in our analysis, we would refer the deformation parameter as being $ \kappa $ instead of $ \eta $. Notice that, under this definition of variables, the large ($ \kappa \rightarrow \infty $) deformation limit in the new variable corresponds to taking the limit, $ \eta \rightarrow 1 $ in terms of the original deformation parameter. },
\begin{eqnarray}
\mathfrak{h}&=&\frac{1+\varrho^{2}}{(1-\kappa^{2}\varrho^{2})},~~\mathfrak{f}=\frac{1}{(1+\varrho^{2})(1-\kappa^{2}\varrho^{2})}\nonumber\\
\tilde{\mathfrak{h}}&=&\frac{\sin^{2}\theta}{(1+\kappa^{2}\cos^{2}\theta)},~~\tilde{\mathfrak{f}}=\frac{1}{(1+\kappa^{2}\cos^{2}\theta)}
\end{eqnarray}
such that the Kalb Ramond two form ($ B $) vanishes during the process of consistent $ 6D $ reduction starting from the original $ AdS_5 \times S^{5}$ solution \cite{Hoare:2014pna}. Notice that, here $ \varphi $, $ \theta $ and $ \phi $ characterize the so called angular coordinates on $ \kappa $-deformed $ S^{3} $. Before we proceed further, it is customary to notice that as there does not seem to be any coordinate mixing between the two subspaces, hence one could decouple the $ S^{3} $ part from the rest of the analysis and perform computations only for the deformed $ AdS_3 $ sector.

Notice that the deformed $ AdS_3 $ sector of the above target space (\ref{E1}) exhibits a non trivial singularity at a finite radial distance, $ \varrho \sim \kappa^{-1} $ \cite{Hoare:2014pna}. As a natural consequence of this, one needs to put the corresponding dual field theory at a finite cutoff surface close enough to $ \varrho \sim\kappa^{-1} $ and imagine the so called \textit{holographic screen} \cite{Kameyama:2014vma}-\cite{Kameyama:2014via} that replaces the usual notion of the boundary in generic AdS/CFT correspondence. In other words, the holographic screen defines a valid physical region, $ 0<\varrho \leq \kappa^{-1} $ within the bulk where holographic computations make sense. 

The first step towards our current analysis would be to perform a series of coordinate transformations in order to express the deformed $ AdS_{3} $ sector of (\ref{E1}) in the so called Poincare coordinates with proper Lorentzian signature. This would eventually lead us towards a Poincare version of the $ \kappa $- deformed $ AdS_3 $ that possesses smooth conformally flat $ AdS_3 $ space-time in the limit of the vanishing ($ \kappa \rightarrow 0 $) deformations. 

To start with, we define the so called \textit{global} coordinates namely,
\begin{eqnarray}
\varrho = \sinh \chi\label{E3}
\end{eqnarray}
which finally leads to,
\begin{eqnarray}
ds^{2}_{AdS_3}=-\left( \frac{\cosh^{2}\chi}{1-\kappa^{2}\sinh^{2}\chi}\right)dt^{2}+\frac{d\chi^{2}}{(1-\kappa^{2}\sinh^{2}\chi)}+\sinh^{2}\chi d\psi^{2}.\label{E4}
\end{eqnarray}

For better understanding of the holographic correspondence, it is customary to express (\ref{E4}) first in the Euclidean signature and then Wick rotate back it to the usual Lorentzian signature. In order to proceed further, we perform a Wick rotation ($ t \rightarrow i\tilde{t} $) along the time axis and define the following map namely,
\begin{eqnarray}
\cosh \chi =\frac{1}{\cos \gamma}\label{E5}
\end{eqnarray}
which yields the following,
\begin{eqnarray}
ds^{2}_{EAdS_3}=\frac{(d\tilde{t}^{2}+d\gamma^{2})}{\cos^{2}\gamma -\kappa^{2}\sin^{2}\gamma}+\frac{\sin^{2}\gamma}{\cos^{2}\gamma}d\psi^{2}.\label{E6}
\end{eqnarray}

Next, we define the following set of coordinate transformations,
\begin{eqnarray}
z=e^{\tilde{t}}\cos \gamma ,~~r=e^{\tilde{t}} \sin \gamma
\label{E7}
\end{eqnarray}
that essentially yields,
\begin{eqnarray}
ds^{2}_{EAdS_3}=\frac{1}{z^{2}}\left( \frac{dz^{2}+dr^{2}}{1-\frac{\kappa^{2}r^{2}}{z^{2}}} + r^{2}d\psi^{2}\right).\label{E8}
\end{eqnarray}

In order to get back to the usual Cartesian coordinates, we define
\begin{eqnarray}
x^{0}=\mathfrak{t}=r \sin \psi ,~~x^{1}=\mathfrak{x}= r \cos \psi
\end{eqnarray}
which eventually leads to the following Euclidean metric \cite{Roychowdhury:2017oqd},
\begin{eqnarray}
ds^{2}_{EAdS_3}=\frac{1}{z^{2}}\left( \frac{dz^{2}+dx^{2}}{1-\frac{\kappa^{2}x^{2}}{z^{2}}} \right)-\frac{\kappa^{2}(\mathfrak{t}d \mathfrak{x}-\mathfrak{x}d \mathfrak{t})^{2}}{z^{2}(1-\frac{\kappa^{2}x^{2}}{z^{2}})}.
\label{E10}
\end{eqnarray}

Finally, performing a second Wick rotation one essentially arrives at the $ \kappa $- deformed Poincare metric of the following form\footnote{Notice that throughout our analysis we set the length scale of the AdS equal to unity. Therefore in our analysis, we essentially work with entities with no mass dimension.},
\begin{eqnarray}
ds^{2}_{LAdS_3}=\frac{1}{z^{2}}\left( \frac{dz^{2}+\eta_{ab}d\mathfrak{x}^{a}d\mathfrak{x}^{b}}{1-\frac{\kappa^{2}x_{L}^{2}}{z^{2}}} \right)+\frac{\kappa^{2}(\mathfrak{t}d \mathfrak{x}-\mathfrak{x}d \mathfrak{t})^{2}}{z^{2}(1-\frac{\kappa^{2}x_L^{2}}{z^{2}})}
\label{E12}
\end{eqnarray}
where the entity, $ x_{L}^{2}= \mathfrak{t}^{2}-\mathfrak{x}^{2}$ denotes the square of the distance of separation between two events at the boundary in the Lorentzian signature, $\eta_{ab}=diag( +1,-1) $. 
However, for the sake of our current analysis, we rewrite (\ref{E12}) in the following way,
\begin{eqnarray}
ds^{2}_{LAdS_3}=\mathcal{Z}(z,x_L)dz^{2}+\mathcal{T}(z,x_L)d\mathfrak{t}^{2}+\mathcal{X}(z,x_L)d\mathfrak{x}^{2}+2\mathcal{K}(z,x_L)d\mathfrak{t}d\mathfrak{x}\label{E14}
\end{eqnarray}
where, the individual metric functions could be formally expressed as,
\begin{eqnarray}
\mathcal{Z}(z,x_L)&=&\frac{1}{z^{2}(1-\frac{\kappa^{2}x_L^{2}}{z^{2}})},~~
\mathcal{T}(z,x_L)=\frac{(1+\kappa^{2}\mathfrak{x}^{2})}{z^{2}\left( 1-\frac{\kappa^{2}x_L^{2}}{z^{2}}\right) }\nonumber\\
\mathcal{X}(z,x_L)&=&\frac{-(1-\kappa^{2}\mathfrak{t}^{2})}{z^{2}\left( 1-\frac{\kappa^{2}x_L^{2}}{z^{2}}\right) },~~
\mathcal{K}(z,x_L)=\frac{-\kappa^{2}\mathfrak{t x}}{z^{2}\left( 1-\frac{\kappa^{2}x_{L}^{2}}{z^{2}}\right) }.\label{E15}
\end{eqnarray}

\section{Retarded correlators}
The purpose of this Section is to explore the low frequency (as well as the corresponding low momentum) behaviour associated with retarded two point correlators by considering fluctuations corresponding to the worldvolume gauge fields associated with the $ N_f $ flavour $ Dp $ brane configurations at the linearised level. In particular, we explore retarded current-current and density-density two point function at zero temperature and finite chemical potential ($ \mu $). The motivation for our present computation rests over the previous novel observations \cite{Karch:2009zz} which indicate that one should be able to find traces of zero sound modes in the pole structure associated with the retarded two point correlators at low frequencies.

In order to compute the density and current correlators at strong coupling, one first needs to expand the \textit{on-shell} DBI action,
\begin{eqnarray}
\mathfrak{S}=-N_f \mathfrak{T}_{p}\int d\tau d^{n}\sigma \sqrt{-det(\mathfrak{G}_{ab}+\mathfrak{F}_{ab})}=N_f \mathfrak{T}_{p}\int d\tau d^{n}\sigma \mathfrak{L}\label{dbi}
\end{eqnarray}
upto quadratic order in the corresponding $ U(1) $ fluctuations (sources) namely\footnote{At this stage, it is noteworthy to mention that the gauge fluctuations are also in the diagonal of the corresponding flavour gauge group $ U(N_f) $. },
\begin{eqnarray}
\mathfrak{A}_{\mu}\rightarrow \mathfrak{A}_{\mu} (z)+\mathfrak{a}_{\mu}(z,\mathfrak{t},\mathfrak{x})
\end{eqnarray}
where, $ \mu =z,\mathfrak{t},\mathfrak{x} $ and $ \mathfrak{a}_{z}=0 $ \cite{Karch:2009zz}. Here, $ \mathfrak{T}_{p} $ is the tension associated with flavour $ Dp $ branes together with $ \tau $ and $ \sigma^{a}(a=1,..,n) $ as being the worldvolume coordinates. Here, $ \mathfrak{G}_{ab}=g_{\mu \nu}\partial_{a}X^{\mu}\partial_{b} X^{\nu}$ is the induced metric on the worldvolume of the $ Dp $ brane and $ \mathfrak{F}_{ab} $ is the $ U(1) $ field strength tensor associated with the corresponding (abelian) worldvolume gauge field.

The DBI action (\ref{dbi}), expanded upto quadratic order in the fluctuations reads\footnote{At this stage, one should notice that in the limit of the vanishing ($ \kappa \rightarrow 0 $) background deformations, the corresponding DBI action (\ref{E44}) boils down to the usual undeformed form \cite{HoyosBadajoz:2010kd} with signature $ (+,-) $.},
\begin{eqnarray}
\mathfrak{S}^{(2)}=\frac{N_{f}\mathfrak{T}_{p}}{2}\int d\mathfrak{t}d\mathfrak{x}dz \left[\frac{\mathcal{T}\mathfrak{a}^{'2}_{\mathfrak{x}}+\mathfrak{f}^{2}_{\mathfrak{tx}}\mathcal{Z}}{\sqrt{\mathcal{D}}} -\frac{\mathcal{N}\mathfrak{a}'_{\mathfrak{x}}\mathfrak{a}'_{\mathfrak{t}}+\mathcal{Q}\mathfrak{a}^{'2}_{\mathfrak{t}}-\mathcal{K}^{2}\mathfrak{A}^{'2}_{\mathfrak{t}}\mathfrak{a}^{'2}_{\mathfrak{x}}}{\mathcal{D}^{3/2}}\right]\label{E44}
\end{eqnarray} 
where, each of the individual entities above (\ref{E44}) could be formally expressed as,
\begin{eqnarray}
\mathcal{D}(z,\Delta x_L)&=&\mathcal{T}|\mathcal{X}|\mathcal{Z}+\mathcal{K}^{2}\mathcal{Z}+|\mathcal{X}|\mathfrak{A}^{'2}_{\mathfrak{t}}\nonumber\\
\mathcal{N}(z,\Delta x_L)&=&2\mathcal{K}\mathcal{Z}(\mathcal{K}^{2}+\mathcal{T}|\mathcal{X}|)\nonumber\\
\mathcal{Q}(z,\Delta x_L)&=& \mathcal{Z}|\mathcal{X}|(\mathcal{K}^{2}+\mathcal{T}|\mathcal{X}|)\nonumber\\
\mathfrak{f}_{\mathfrak{tx}}&=&\partial_{\mathfrak{t}}\mathfrak{a}_{\mathfrak{x}}-\partial_{\mathfrak{x}}\mathfrak{a}_{\mathfrak{t}}.
\end{eqnarray}
Notice that, here, $ \Delta x_L =\sqrt{\Delta T^{2}-\Delta L^{2}}\geq 0 $ corresponds to some \textit{fixed} time like separation associated with the dual field theory living on the hypersurface (at a fixed radial distance, $ z = z_0 $) that is infinitesimally closed to the holographic screen \cite{Kameyama:2014vma}-\cite{Kameyama:2014via} mentioned earlier. It is noteworthy to mention that, here $ \Delta x_L ^{2} $ corresponds to the value of the function $ x^{2}_{L} $ appearing in (\ref{E12}) near the holographic screen, $ z=z_0 \sim z_{B} $. Here $ z_B $ is the location of the holographic screen that might be regarded as the UV cut-off ($ \epsilon_{UV}\equiv \epsilon $) of the theory.

Our next task would be to solve these fluctuations in the momentum space. In order to do this, we consider the following Fourier transform,
\begin{eqnarray}
\mathfrak{a}_{\mu}(z,\mathfrak{t},\mathfrak{x})=\int \frac{d\mathfrak{w}d\mathfrak{q}}{(2\pi)^{2}}e^{-i \mathfrak{w}\mathfrak{t}+i\mathfrak{q}\mathfrak{x}}\mathfrak{a}_{\mu}(z,\mathfrak{w},\mathfrak{q}).\label{E46}
\end{eqnarray} 

Substituting (\ref{E46}) back into (\ref{E44}), we arrive at the following set of equations of motion,
\begin{eqnarray}
\partial_{z}\left( \frac{\mathcal{Q}\mathfrak{a}'_{\mathfrak{t}}}{\mathcal{D}^{3/2}}\right) +\frac{1}{2}\partial_{z}\left( \frac{\mathcal{N}\mathfrak{a}'_{\mathfrak{x}}}{\mathcal{D}^{3/2}}\right)-(\mathfrak{w q}\mathfrak{a}_{\mathfrak{x}}+\mathfrak{q}^{2}\mathfrak{a}_{\mathfrak{t}})\frac{\mathcal{Z}}{\sqrt{\mathcal{D}}}&=&0\nonumber\\
\partial_{z}\left( \frac{\mathcal{T}\mathfrak{a}'_{\mathfrak{x}}}{\sqrt{\mathcal{D}}}\right) -\frac{1}{2}\partial_{z}\left( \frac{\mathcal{N}\mathfrak{a}'_{\mathfrak{t}}}{\mathcal{D}^{3/2}}\right)+\partial_{z}\left( \frac{\mathcal{K}^{2}\mathfrak{A}^{'2}_{\mathfrak{t}}\mathfrak{a}'_{\mathfrak{x}}}{\mathcal{D}^{3/2}}\right) +(\mathfrak{w q}\mathfrak{a}_{\mathfrak{t}}+\mathfrak{w}^{2}\mathfrak{a}_{\mathfrak{x}})\frac{\mathcal{Z}}{\sqrt{\mathcal{D}}}&=&0.\label{E47}
 \end{eqnarray}

Clearly, compared with the earlier results \cite{HoyosBadajoz:2010kd}, here we observe the emergence of additional structures in (\ref{E47}) those are proportional to functions like $ \mathcal{N}(z,\Delta x_L) $ or $ \mathcal{K} (z,\Delta x_L) $ which are therefore purely contributions due to background $ \kappa $- deformations.
However, as noticed earlier \cite{Karch:2009zz}, there also exists another (constraint) equation corresponding to $\mathfrak{a}_{z}$ (subjected to the gauge choice, $\mathfrak{a}_{z}=0$) which is obtained by varying the DBI action (\ref{dbi}),
\begin{eqnarray}
\mathcal{D}^{3/2}\partial_{\mathfrak{x}}\left( \frac{\mathcal{T}}{\sqrt{\mathcal{D}}}\right) \mathfrak{a}'_{\mathfrak{x}}+i\mathfrak{q}\mathcal{T}\mathcal{D}\mathfrak{a}'_{\mathfrak{x}}+\frac{\mathcal{K}}{2}\mathfrak{A}'_{\mathfrak{t}}\partial_{\mathfrak{x}}\mathcal{D}^{(1)}-\mathcal{D}^{3/2}\partial_{\mathfrak{x}}\left( \frac{\mathcal{K}}{\sqrt{\mathcal{D}}}\right)\mathfrak{a}'_{\mathfrak{t}}
-i\mathfrak{q}\mathcal{K D}\mathfrak{a}'_{\mathfrak{t}}\nonumber\\
-\mathcal{D}^{3/2}\partial_{\mathfrak{t}}\left( \frac{\mathcal{K}}{\sqrt{\mathcal{D}}}\right)\mathfrak{a}'_{\mathfrak{x}}+i\mathfrak{w}\mathcal{KD}\mathfrak{a}'_{\mathfrak{x}}+\frac{|\mathcal{\chi }|}{2}\mathfrak{A}'_{\mathfrak{t}}\partial_{\mathfrak{t}}\mathcal{D}^{(1)}-\mathcal{D}^{3/2}\partial_{\mathfrak{t}}\left( \frac{|\mathcal{\chi }|}{\sqrt{\mathcal{D}}}\right)\mathfrak{a}'_{\mathfrak{t}}+i\mathfrak{w}\mathcal{D}|\chi |\mathfrak{a}'_{\mathfrak{t}}=0\label{E48}
\end{eqnarray}
where, the entity,
\begin{eqnarray}
 \mathcal{D}^{(1)}=2 \mathcal{K}\mathfrak{A}'_{\mathfrak{t}}\mathfrak{a}'_{\mathfrak{x}}+2 | \chi |\mathfrak{A}'_{\mathfrak{t}}\mathfrak{a}'_{\mathfrak{t}}
\end{eqnarray}
contains terms only first order in the fluctuations in the DBI action (\ref{dbi}).

For the purpose of our present analysis, we rewrite (\ref{E48}) as,
\begin{eqnarray}
\mathfrak{H}(z,x^{a})\mathfrak{a}'_{\mathfrak{x}}-\mathfrak{Q}(z,x^{a})\mathfrak{a}'_{\mathfrak{t}}=0 \label{E49}
\end{eqnarray}
where, the individual functions above in (\ref{E49}) could be formally expressed as,
\begin{eqnarray}
\mathfrak{H}(z,x^{a})&=&\mathcal{D}^{3/2}\left( \partial_{\mathfrak{x}}\left( \frac{\mathcal{T}}{\sqrt{\mathcal{D}}}\right)-\partial_{\mathfrak{t}}\left( \frac{\mathcal{K}}{\sqrt{\mathcal{D}}}\right)\right) +i\mathcal{D}(\mathfrak{q}\mathcal{T}+\mathfrak{w}\mathcal{K})+i\mathfrak{A}^{'2}_{\mathfrak{t}}\mathcal{K}(\mathfrak{q}\mathcal{K}-\mathfrak{w}|\chi |)\nonumber\\
\mathfrak{Q}(z,x^{a})&=&\mathcal{D}^{3/2}\left( \partial_{\mathfrak{x}}\left( \frac{\mathcal{K}}{\sqrt{\mathcal{D}}}\right)+\partial_{\mathfrak{t}}\left( \frac{|\chi |}{\sqrt{\mathcal{D}}}\right)\right) +i\mathcal{D}(\mathfrak{q}\mathcal{K}-\mathfrak{w}|\chi |)-i\mathfrak{A}^{'2}_{\mathfrak{t}}|\chi |(\mathfrak{q}\mathcal{K}-\mathfrak{w}|\chi |).\nonumber\\
\end{eqnarray}

Using (\ref{E49}), one could re-express (\ref{E47}) as,
\begin{eqnarray}
\mathfrak{a}'_{\mathfrak{t}}=\frac{\mathfrak{C}^{(1)}}{\Theta (\mathfrak{w},\mathfrak{q},z,x^{a})}
\end{eqnarray}
where, $ \mathfrak{C}^{(1)} $ is some integration constant and the function at the denominator could be formally expressed as,
\begin{eqnarray}
\Theta (\mathfrak{w},\mathfrak{q},z,x^{a})=\mathcal{Q}+\frac{\mathcal{N}\mathfrak{Q}}{2\mathfrak{H}}+\frac{\mathfrak{q}}{\mathfrak{w}}\frac{\mathcal{T D}\mathfrak{Q}}{\mathfrak{H}}-\frac{\mathfrak{q}\mathcal{N}}{2\mathfrak{w}}+\frac{\mathfrak{q}\mathfrak{Q}}{\mathfrak{w}\mathfrak{H}}\mathcal{K}^{2}\mathfrak{A}^{'2}_{\mathfrak{t}}.
\end{eqnarray}

Notice that, the set of equations (\ref{E47})-(\ref{E48}) possesses the residual gauge symmtery of the following form,
\begin{eqnarray}
\mathfrak{a}_{\mathfrak{t}}\rightarrow \mathfrak{a}_{\mathfrak{t}} - \mathfrak{w}\xi (\mathfrak{t},\mathfrak{x})\nonumber\\
\mathfrak{a}_{\mathfrak{x}}\rightarrow \mathfrak{a}_{\mathfrak{x}} + \mathfrak{q}\xi (\mathfrak{t},\mathfrak{x})
\end{eqnarray}
which thereby implies that only physical degrees of freedom of the system are those which are gauge invariant. Therefore, in our analysis, instead of solving the equation for $ \mathfrak{a}_{\mathfrak{t}} $, we would rather solve the following gauge invariant combination namely the electric field\cite{Karch:2009zz},
\begin{eqnarray}
\mathfrak{E}=\mathfrak{w}\mathfrak{a}_{\mathfrak{x}}+\mathfrak{q}\mathfrak{a}_{\mathfrak{t}}.\label{E53}
\end{eqnarray} 

Using (\ref{E53}) and the constrained equation (\ref{E49}) we finally obtain,
\begin{eqnarray}
\mathfrak{E}'=\frac{\mathfrak{w}\mathfrak{C}^{(1)}}{\mathfrak{H}\Theta}\left(\mathfrak{Q}+\frac{\mathfrak{q}}{\mathfrak{w}}\mathfrak{H} \right).\label{E54}
\end{eqnarray}

Substituting (\ref{E53}) into (\ref{E44}) and using the Fourier transform (\ref{E46}), the corresponding DBI action 
(in the frequency space) finally turns out to be,
\begin{eqnarray}
\mathfrak{S}^{(2)}|_{z\sim \epsilon}\approx\frac{N_{f}\mathfrak{T}_{p}}{2}\int _{z\sim \epsilon}dz d\mathfrak{w}d\mathfrak{q}~\mathfrak{I}(\mathfrak{w},\mathfrak{q},z)
\end{eqnarray}
where, the integrand could be formally expressed as,
\begin{eqnarray}
\mathfrak{I}(\mathfrak{w},\mathfrak{q},z)&=&\frac{1}{\mathfrak{w}^{2}}\left(\langle\mathfrak{Q}\rangle+\frac{\mathfrak{q}\langle\mathfrak{H}\rangle}{\mathfrak{w}} \right)^{-2}\left(\frac{\langle\mathfrak{Q}^{2}\rangle\langle\mathcal{T}\rangle}{\sqrt{\langle\mathcal{D}\rangle}}-\frac{(\langle\mathcal{N}\mathfrak{Q}\mathfrak{H}\rangle +\langle \mathcal{Q}\mathfrak{H}^{2}\rangle -\langle \mathcal{K}^{2}\mathfrak{A}^{'2}_{\mathfrak{t}}\mathfrak{Q}^{2}\rangle)}{\langle \mathcal{D}\rangle^{3/2}} \right) \mathfrak{E}^{'2} 
-\frac{\langle \mathcal{Z}\rangle} {\sqrt{\langle \mathcal{D}\rangle}}\mathfrak{E}^{2}\nonumber\\
&=&\mathcal{B}(z, \kappa ) \mathfrak{E}^{'2}-\frac{\langle \mathcal{Z}\rangle} {\sqrt{\langle \mathcal{D}\rangle}}\mathfrak{E}^{2}
\end{eqnarray}
where, by performing the above computation we always assume that we are close to the UV cut-off, namely $ z\sim \epsilon \simeq z_B $ where we consider
a fixed time like separation, $ \Delta x_L =\sqrt{\Delta T^{2}-\Delta L^{2}}\geq 0 $. The following analysis is valid only when one could average over temporal as well as spatial dependencies appearing in the metric functions (\ref{E15}) by their respective values corresponding to ($ \Delta T, \Delta L $) namely,
\begin{eqnarray}
\mathcal{T}(\mathfrak{x}^{a},z)|_{z\sim \epsilon}\sim \langle \mathcal{T}(z)\rangle_{z\sim \epsilon}\simeq \mathcal{T}(\Delta L,  \epsilon)\label{A}
\end{eqnarray}
and so on. This is also equivalent of considering the fact that the metric functions near the boundary is a slowly varying $ \left(|\frac{\delta g_{ab}}{\delta \mathfrak{x}^{a}}|\ll 1 \right)  $ function of the boundary coordinates so that we could replace them by some suitable averages of the above form\footnote{Under such an approximation, one could in fact get rid off the spatial integrals appearing in (\ref{E44}) by using the usual definition of Fourier transform of delta function in the momentum space.} (\ref{A}). 

Finally, by performing the integration by parts, the boundary action turns out to be,
\begin{eqnarray}
\mathfrak{S}^{(2)}_{B}|_{z\sim \epsilon}=-\frac{N_{f}\mathfrak{T}_{p}\epsilon}{2}\int d\mathfrak{w}d\mathfrak{q}\mathcal{B}(\mathfrak{w},\mathfrak{q},\kappa)\mathfrak{E}(\epsilon)\mathfrak{E}'(\epsilon)\label{E58}
\end{eqnarray}
where, the function $ \mathcal{B}(\epsilon ,\kappa) $ could be formally expressed as,
\begin{eqnarray}
\mathcal{B}(\mathfrak{w},\mathfrak{q},\kappa)\approx \frac{ \sqrt{1-2 \kappa ^4 \Delta L^4}}{(\mathfrak{q}^{2}+\mathfrak{w}^{2}+(\mathfrak{q}^{2}-\mathfrak{w}^{2}-2 \mathfrak{q w})\kappa^{2}\Delta L^{2})}.
\end{eqnarray}

Given the on-shell action (\ref{E58}), one can immediately write down the retarded two point correlators by taking functional derivatives \cite{HoyosBadajoz:2010kd},
\begin{eqnarray}
\mathcal{G}_{\mathfrak{tt}}^{R}&=&\mathfrak{q}^{2}\mathcal{B}(\mathfrak{w},\mathfrak{q},\kappa) \Pi (\mathfrak{w},\mathfrak{q},\kappa)\nonumber\\
\mathcal{G}_{\mathfrak{tx}}^{R}&=&\mathfrak{q}\mathfrak{w}\mathcal{B}(\mathfrak{w},\mathfrak{q},\kappa) \Pi (\mathfrak{w},\mathfrak{q},\kappa)\nonumber\\
\mathcal{G}_{\mathfrak{xx}}^{R}&=&\mathfrak{w}^{2}\mathcal{B}(\mathfrak{w},\mathfrak{q},\kappa) \Pi (\mathfrak{w},\mathfrak{q},\kappa)\label{E60}
\end{eqnarray}
where, we identify,
\begin{eqnarray}
\Pi (\mathfrak{w},\mathfrak{q},\kappa)=\frac{\delta^{2}\mathfrak{S}^{(2)}_{B}}{\delta \mathfrak{E}^{2}(\epsilon)}.
\end{eqnarray}

The solution to the Eq.(\ref{E54}) contains to unkown integration constants ($ \mathfrak{C}^{(1)} $ and $ \mathfrak{C}^{(2)} $) which are to be determined from some physical considerations. Notice that Eq.(\ref{E54}) is valid for all $ z $.  Since we are interested in computing retarded correlators (which are related to the response parameters of the system) therefore one must take into account the effect of attenuation and/or the dissipation into the system and which is related to the fact that the low frequency ($ \mathfrak{w}z^{n} \ll 1 $) as well as low momentum ($ \mathfrak{q}z^{n} \ll 1 $) modes (collectively the \textit{hydrodynamic} modes) must satisfy in-going wave boundary conditions in the interior of the bulk\footnote{Here, $ n (\geq 9)$ is some sufficiently large positive integer.} \cite{HoyosBadajoz:2010kd}. Therefore, the fluctuations that we finally solve for (near the boundary) must be consistent with the in-going wave boundary condition in the bulk. This information is hidden in the above integration constants ($ \mathfrak{C}^{(1)} $ and $ \mathfrak{C}^{(2)} $) that finally appear in the retarded correlator (\ref{E64}). In other words, the retarded correlator we compute is a physical one and is consistent with in-going wave boundary condition. In order to determine these constants we need a matching criteria which is obtained (i) first by expanding (\ref{E54}) into large $ z $ and then taking a small frequency expansion and (ii) to solve (\ref{E54}) for small frequency first and then expanding into large $ z $.

However, before getting into that, it is customary note down the low frequency solution close to the boundary, 
\begin{eqnarray}
\mathfrak{E} (z)\simeq \mathfrak{C}^{(2)}+\mathfrak{C}^{(1)}(\alpha \mathfrak{w} +\beta \mathfrak{w q}+\gamma \frac{\mathfrak{w}^{2}}{\mathfrak{q}}+\zeta \mathfrak{w}^{2})z\label{E62}
\end{eqnarray}
along with the properly defined coefficients of the following form,
\begin{eqnarray}
\alpha &=& \frac{\sqrt{\Delta L^2}  \left(\kappa ^4 \Delta L^4-\kappa ^2 \Delta L^2-1\right)}{\kappa ^2 \Delta L^3 \left(1-2 \kappa ^4 \Delta L^4\right)^2}\nonumber\\
\beta &=&\frac{i  \sqrt{\Delta L^2}  \left(\kappa ^2 \Delta L^2+1\right)}{4 \kappa ^8 \Delta L^8-2 \kappa ^4 \Delta L^4}\nonumber\\
\gamma &=&\frac{ \left(\kappa ^2 \Delta L^2-1\right) \left(\kappa ^2 \Delta L^2+1\right)^2}{\kappa ^4  \Delta L^4 \left(1-2 \kappa ^4 \Delta L^4\right)^2}\nonumber\\
\zeta &=&\frac{i  \left(-2 \kappa ^4 \Delta L^4+\kappa ^2\Delta L^2+1\right)}{2 \kappa ^6 \Delta L^5 \left(2 \kappa ^4 \Delta L^4-1\right)}.
\end{eqnarray}

Substituting (\ref{E62}) into (\ref{E58}) we find,
\begin{eqnarray}
\mathfrak{S}^{(2)}_{B}|_{z\sim \epsilon}=-\frac{N_{f}\mathfrak{T}_{p}}{2}\int d\mathfrak{w}d\mathfrak{q}\mathfrak{C}^{(1)}\mathfrak{C}^{(2)}\mathcal{B}(\mathfrak{w},\mathfrak{q},\kappa)(\alpha \mathfrak{w} +\beta \mathfrak{w q}+\gamma \frac{\mathfrak{w}^{2}}{\mathfrak{q}}+\zeta \mathfrak{w}^{2})\label{E64}
\end{eqnarray}

Our next task would be to determine the constants $ \mathfrak{C}^{(1)} $ and $ \mathfrak{C}^{(2)} $ following the physical arguments metioned above. The first step would to solve fluctuations first in the limit $ z \rightarrow \infty $ and then consider small frequency ($ \mathfrak{w}z^{n} \ll 1 $) as well as small momentum ($ \mathfrak{q}z^{n} \ll 1 $) expansion while keeping the ratio, $ \frac{\mathfrak{w}}{\mathfrak{q}} $ fixed. Our analysis reveals that in the large $ z $ limit, the solution (\ref{E54}) indeed behaves like an ingoing wave of the following form\cite{HoyosBadajoz:2010kd},
\begin{eqnarray}
\mathfrak{E}(z) \sim e^{i (\mathfrak{w}\varsigma -\hat{q}\mathfrak{q}) z^{9}/9}
\end{eqnarray}
such that the entity, $ \varsigma  $ could be formally expressed as,
\begin{eqnarray}
\varsigma =\frac{ (2\tilde{\omega} -\tilde{q})}{2 \kappa ^2 \Delta \mathfrak{t}^2}-\frac{\tilde{q}}{2}+\frac{2  \Delta \mathfrak{x} \left(2 \tilde{\omega}\Delta \mathfrak{x}+\Delta \mathfrak{t}(\hat{\omega}+i)\right)}{\Delta\mathfrak{t}^2}
\end{eqnarray}
where, we set, $ \tilde{\omega}= |\mathfrak{w}\Delta \mathfrak{x}|\ll 1$, $ \tilde{q}= |\mathfrak{q}\Delta \mathfrak{t}|\ll 1$ , $ \hat{\omega}= |\mathfrak{w}\Delta \mathfrak{t}|\ll 1$ and, $ \hat{q}= |\mathfrak{q}\Delta \mathfrak{x}| \ll 1$.

Next, we would like to take the small frequency ($ \mathfrak{w}z^{n} \ll 1 $) as well as the small momentum ($ \mathfrak{q}z^{n} \ll 1 $) limit of the large $ z $ solution \cite{HoyosBadajoz:2010kd} corresponding to (\ref{E54}) which finally yields,
\begin{eqnarray}
\mathfrak{E}(z)\approx \mathfrak{C}^{(3)}+\frac{\mathfrak{C}^{(3)}\mathfrak{w} z^9 \left(1-\kappa ^2 \Delta\mathfrak{t}^2\right) \left(1-\kappa^{2}\Delta\mathfrak{x}^{2}\right)}{9\kappa ^2 \Delta\mathfrak{t}\Delta\mathfrak{x} \left(2 \kappa ^4 \Delta\mathfrak{t}^2 \Delta\mathfrak{x}^2+\kappa ^2 \Delta \mathfrak{s}^{2}-1\right) \left(\kappa ^2 \Delta\mathfrak{t}^2 \left(2 \kappa ^2\Delta\mathfrak{x}^2-1\right)+1\right)}\nonumber\\
+\frac{i \mathfrak{C}^{(3)}\mathfrak{w q} z^9}{9\Delta\mathfrak{t}^3 \left(4 \kappa ^6 \Delta\mathfrak{x}^2-2 \kappa ^4\right)+2 \kappa ^2 \Delta\mathfrak{t}}\label{E67}
\end{eqnarray}
where, $ \Delta \mathfrak{s}^{2}=\Delta\mathfrak{t}^{2}- \Delta\mathfrak{x}^{2}\geq 0$ stands for some fixed time like interval corresponding to some fixed IR ($ z\rightarrow \infty $) hyper-surface at the deep interior of the bulk. 

Following the approach of \cite{HoyosBadajoz:2010kd}, our next task would be first to obtain the general solution (valid for arbitrary values of $ z $) corresponding to (\ref{E54}) in the regime of small frequency ($ \mathfrak{w}z^{n} \ll 1 $) as well as small momentum ($ \mathfrak{q}z^{n} \ll 1 $) and then consider the large $ z $ limit of this solution which finally yields\footnote{At this stage, it is indeed worthwhile to point out that in order to obtain solution (\ref{E68}), we had taken care of the fact that in the limit of small frequency and momentum and for generic background ($ \kappa >0 $) deformations it is in-fact quite reasonable to set the limit , $ \mathfrak{w}\kappa^{2}\ll 1 $ as well as, $\mathfrak{q}\kappa^{2}\ll 1 $ .},
\begin{eqnarray}
\mathfrak{E}(z)\simeq\mathfrak{C}^{(2)}+\mathfrak{C}^{(1)}\left(\frac{i \mathfrak{w}  z^9 \left(\mathfrak{q}+2 i \kappa ^2 \Delta \mathfrak{x}\right)\Delta \mathfrak{t}}{36 \kappa ^2 \Delta \mathfrak{x}^2} \right) +\mathcal{O}\left( \frac{\mathfrak{w}z^{n}}{z^{m}}\right) \label{E68}
\end{eqnarray}
where, $ n=9 $ and $ m \geq 2 $ are two positive integers.

Comparing (\ref{E67}) and (\ref{E68}), we finally obtain,
\begin{eqnarray}
\mathfrak{C}^{(1)}=\mathfrak{C}^{(2)}\left(\frac{\mathfrak{q}\Gamma^{(2)}-\Gamma^{(1)}}{\mathfrak{q}\Gamma^{(3)}+\Gamma^{(4)}} \right) \label{E69}
\end{eqnarray}
where, the coefficients $ \Gamma $ could be formally expressed as,
\begin{eqnarray}
\Gamma^{(1)}& = &\frac{i\left(1-\kappa ^2 \Delta\mathfrak{t}^2\right) \left(1-\kappa^{2}\Delta\mathfrak{x}^{2}\right)}{(9\kappa ^2 \Delta\mathfrak{t}\Delta\mathfrak{x} \left(2 \kappa ^4 \Delta\mathfrak{t}^2 \Delta\mathfrak{x}^2+\kappa ^2 \Delta \mathfrak{s}^{2}-1\right) \left(\kappa ^2 \Delta\mathfrak{t}^2 \left(2 \kappa ^2\Delta\mathfrak{x}^2-1\right)+1\right))}\nonumber\\
\Gamma^{(2)}& = &\frac{1}{(9\Delta\mathfrak{t}^3 \left(4 \kappa ^6 \Delta\mathfrak{x}^2-2 \kappa ^4\right)+2 \kappa ^2 \Delta\mathfrak{t})}\nonumber\\
\Gamma^{(3)}& = &\frac{1}{36 \kappa^{2}\Delta \mathfrak{x}^{2}},~~\Gamma^{(4)}=\frac{i \Delta \mathfrak{t}}{18\Delta \mathfrak{x}}.
\end{eqnarray}

Using (\ref{E64}) and (\ref{E69}), the low frequency (as well as the low momentum) behaviour associated with the retarded two point correlator finally turns out to be,
\begin{eqnarray}
\tilde{\Pi}(\mathfrak{w},\mathfrak{q},\kappa)&=&\mathcal{B}(\mathfrak{w},\mathfrak{q},\kappa)\Pi (\mathfrak{w},\mathfrak{q},\kappa)\nonumber\\
&=& \frac{\mathcal{C}\mathfrak{w}}{\mathfrak{c}_{1}\mathfrak{q}^{2}-\mathfrak{c}_{2}\mathfrak{w}^{2}-\frac{\gamma \mathfrak{c}_1}{\alpha}\mathfrak{wq}-\Sigma (\mathfrak{w},\mathfrak{q},\kappa)}\label{E71}
\end{eqnarray}
where, apart from some overall numerical (real) pre-factor ($ \mathcal{C} $), one could formally express the other entities as,
\begin{eqnarray*}
\mathfrak{c}_{1}=\frac{1}{\sqrt{1-2\kappa^{4}\Delta L^{4}}},~~\mathfrak{c}_{2}=\mathfrak{c}_{1}\left( \frac{\gamma \mathfrak{n}}{\alpha}-1\right) ,~~\mathfrak{n}=\frac{\mathfrak{w}}{\mathfrak{q}},~~\chi =\frac{\Gamma^{(3)}}{\Gamma^{(4)}}+\frac{\Gamma^{(2)}}{\Gamma^{(1)}}
\end{eqnarray*}
\begin{eqnarray}
\Sigma (\mathfrak{w},\mathfrak{q},\kappa) = \frac{\mathfrak{c}_{1}}{\alpha} \left(\zeta +\gamma \chi \right) \mathfrak{w}^{3}+\mathfrak{c}_{1}\left(\frac{\beta}{\alpha} -\chi \right) \mathfrak{q}^{3}+\mathfrak{c}_{1}\left(\frac{\beta}{\alpha} -\chi \right) \mathfrak{w}^{2}\mathfrak{q}+\frac{\mathfrak{c}_{1}}{\alpha}\left( \zeta +\gamma \chi\right)  \mathfrak{w}\mathfrak{q}^{2}+\mathcal{O}(4).
\end{eqnarray}

The above two point function (\ref{E71}) is the main result of this paper. It clearly indicates, that at low frequencies the quadratic order fluctuations, $ \mathfrak{w}^{2} $ always dominates over the self energy correction, $ \Sigma $. Therefore, at low frequencies, it is the quadratic order term in the fluctuations that controls the behaviour of the retarded two point correlator (\ref{E71}).

In order to check the existence of zero sound, one first needs to explore the pole of the retarded correlator (\ref{E71}) which amounts to set,
\begin{eqnarray}
\mathfrak{c}_{1}\mathfrak{q}^{2}-\mathfrak{c}_{2}\mathfrak{w}^{2}-\frac{\gamma \mathfrak{c}_1}{\alpha}\mathfrak{wq}-\Sigma (\mathfrak{w},\mathfrak{q},\kappa)=0.\label{E73}
\end{eqnarray}

The above equation (\ref{E73}) could be easily solved in the domain of low frequency regime as a perturbative expansion in the small momentum which finally yields,
\begin{eqnarray}
\mathfrak{w}(\mathfrak{q})\simeq  \mathfrak{d}^{(1)}\mathfrak{q}+\mathfrak{d}^{(2)}\mathfrak{q}^{2}+\mathcal{O}(\mathfrak{q}^{3})\label{E75}
\end{eqnarray}
where, the entities above could be formally expressed as,
\begin{eqnarray}
 \mathfrak{d}^{(1)}&=-&\frac{\gamma \mathfrak{c}_{1}}{2\alpha \mathfrak{c}_{2}}\left( 1\mp \sqrt{1+\frac{4\alpha^{2}\mathfrak{c}_{2}}{\gamma^{2}\mathfrak{c}_{1}}}\right)\nonumber\\
 \mathfrak{d}^{(2)}&=&-\frac{\mathfrak{c}_{1}}{2\alpha \mathfrak{c}_{2}}\left(\zeta +\gamma \chi + \frac{\frac{2\alpha^{2}}{\gamma}\left(\frac{\beta}{\alpha}-\chi \right)+(\zeta +\gamma \chi)}{\sqrt{1+\frac{4\alpha^{2}\mathfrak{c}_{2}}{\gamma^{2}\mathfrak{c}_{1}}}}\right)+\frac{\gamma \mathfrak{c}_{1}^{2}}{2\alpha \mathfrak{c}_{2}^{2}} \left(\frac{\beta}{\alpha}-\chi \right)  \left( 1\mp \sqrt{1+\frac{4\alpha^{2}\mathfrak{c}_{2}}{\gamma^{2}\mathfrak{c}_{1}}}\right).\nonumber\\
\end{eqnarray}
Before we proceed further, a number of crucial observations are in order. First of all, it is worth emphasizing that the above dispersion relation (\ref{E75}) is exact in the background deformations ($ \kappa $) and is therefore valid for any (non-singular) value of $ \kappa(>0) $. The second observation is that the coefficient ($  \mathfrak{d}^{(1)} $) associated with the linear momentum term is purely real whereas on the other hand, the coefficient ($  \mathfrak{d}^{(2)} $) associated with the quadratic momentum term is purely imaginary. This therefore clearly suggests the existence of the so called holographic \textit{zero sound} speed,
\begin{eqnarray}
\upsilon_{0}=\mathfrak{d}^{(1)}\label{e79}
\end{eqnarray}
in the spectrum as the imaginary part falls at a rate faster than that of the real part in the limit of low momentum. In summary, the pole structure associated with the retarded current correlators is quite similar to that what has been observed in previous holographic analysis performed in higher ($ D>2 $) dimensions \cite{Karch:2009zz},\cite{HoyosBadajoz:2010kd}. 
\section{Summary and final remarks}
We conclude our analysis with a brief summary of the analysis performed along with mentioning some of its possible future extensions. In this paper, we explore the low energy behaviour associated with retarded two point correlators for certain classes of ($ 1+1 $)D strongly correlated \textit{quantum liquids} those are dual to classical (bosonic) $ \eta $- deformed $ AdS_3 $ spacetime in ($ 2+1 $)D \cite{Hoare:2014pna}. The corresponding pole structure associated with these retarded correlators reveals an astonishing fact, namely the existence of zero sound like excitations as observed in the context of ordinary LFL theory in higher ($ D>2 $) dimensions. In other words, the quantum liquid (dual to $ \eta $- deformed $ AdS_3 $) behaves like ordinary LFL in ($ 1+1 $)D which is indeed quite unusual as well as surprising from the point of view of ($ 1+1 $)D theory \cite{Hung:2009qk}. Therefore it is actually through \textit{holography} where we have been able to discover some non trivial LFL like phase even in ($ 1+1 $)D. One of the reasons why we observe LFL like phase in ($ 1+1 $)D might have its origin in the so called $ \eta $- deformations. However, this issue does not seem to be resolved at the moment and thereby definitely merits further investigation along this particular direction. 
\\ \\ 
{\bf {Acknowledgements :}}
This work was supported through the Newton-Bhahba Fund. The author would like to acknowledge the Royal Society UK and the Science and Engineering Research Board India (SERB) for financial assistance.\\ 

\end{document}